\documentclass[english]{amsart}
\usepackage[T1]{fontenc}
\usepackage[latin9]{inputenc}

\usepackage{babel}
\usepackage{amsmath, amssymb}

\newcommand{\N}{\mathbb N}

\begin{document}
\title{Remarks on a new possible discretization scheme for gauge theories}
\author{Jean-Pierre Magnot}

\maketitle

\begin{abstract}
We propose here a new discretization method for a class continuum gauge theories which action functionnals are polynomials of the curvature. Based on the notion of holonomy, this discretization procedure appears gauge-invariant for discretized analogs of Yang-Mills theories, and hence gauge-fixing is fully rigorous for these discretized action functionnals. Heuristic parts are forwarded to the quantization procedure via Feynman integrals and the meaning of the heuristic infinite dimensional Lebesgue integral is questionned.
\end{abstract}
\section*{Introduction}
The problem of Gauge fixing in gauge theories is of funcdamental importance for explicit calculations, before or after the quatization procedure, see e.g. \cite{Re1997,AZ1990,Hah2004,Seng2011,Lim2012,HCR2015} for a non-exhaustive list of references.  In the quantized approach, one way to define the Feynman integral over the space of connections is the use of a finite element method for discretizing connections, inspired by \cite{Wh}, see e.g. \cite{AHM,AZ1990,SSSA}. This approach works quite well for abelian gauge theories, but a well-chosen gauge-fixing actually produces with many difficulties some explicit results. Moreover, the invariance under gauge fixing is actually partially justified with heuristic arguments.

The present approach of discretization is based on the following paradigm: the convergence of the discretization \textbf{ for a fixed connection } must be preserved, as well as gauge covariance of the curvature in the discretization scheme. In section \ref{1}, a method for discretizing a connection on a principal bundle over a smooth manifold by a triangulation or a cubification is proposed, even if the manifold or the principal bundles is not trivial. The key tool is no longer differential forms as in \cite{Wh} but the holonomy of the connection. These discretizations are used to define the new discrete analogs of connections and curvature in section \ref{2}. It appears that curvature is covariant with the action of the gauge group, which implies that the discretization procedure in section \ref{1} is less dependent on the choice made arbitrarily than it may appear in a first approach. By the way, some of the discretized analogs obtained in section \ref{3} are gauge-invariant. In particular, this is the case for 2D and 4D-Yang-Mills analogs.  
 
Therefore in the proposed approach, the ``natural'' continuum limit of the quantized model does not involve cylindrical functions on a vector space of connection forms equipped with a Lebesgue measure, but cylindrical functions defined on products of unimodular groups aquipped with their Haar measure. This is discussed in section \ref{2'}. Even if there exists an obvious conceptual link between Haar measure and Lebesgue measure, the two
cylindrical approximations that are produced are not a priori equivalent.  We conclude this presentation conjecturing that the gauge anomalies 
which can appear in classical discretized models may find an expression in this measure defect.  
\section{A new discretization scheme of connections}\label{1}
Let $\pi : P \rightarrow M$ be a principal bundle of connected Riemannian base, with structure group $G,$ equipped with a prescribed triangulation or cubification $\tau.$ The canonical maps induced by $\pi$ on relevant objects will be also noted by $\pi$ in the sequel when it carries no ambiguity. The nodes of this triangulation or cubification are assumed indexed by $\mathbb{N},$ noted by $(s_n)_\mathbb{N}$ (the manifold $M$ can be non compact). Recall that, for a fixed index $i_1,...i_n,$ $St(s_{i_1},...s_{i_n})$ is the domain described by the simplexes or the cubes with nodes $s_i.$  We note by $\mathcal{C}$ the space of connections on $P.$ Let $\theta \in \mathcal{C}.$ Fixing $s_0$ as a basepoint  and $p_0 \in \pi^{-1}(s_0),$ 
\begin{enumerate}
	\item Let $$j = \min \left\{ i \in \N^* | s_i \in St(s_0) \right\}.$$ We define $g_{0,j}=1$  and $p_j$ the endpoint of the horizontal path over $[s_0,s_j]$ with starting point $p_0.$ Let $I_2 = \{0;j\}.$
	\item Assume that $I_n$ exists, and that, 
	\begin{itemize}
		\item $\forall i \in I_n, $ we have constructed $p_i \in \pi^{-1}(s_i)$ and 
		\item $\forall (i,j) \in I_n^2,$ with $i<j$, $g_{i,j}$ is the holonomy of $[s_i,s_j]$, starting at $p_i,$ i.e.  $p_j. g_{i,j}$ is the endpoint of the horizontal path over $[s_i,s_j]$ with starting point $p_i.$ 
	\end{itemize}
Let $$j = \min \left\{ i \in \N- I_n | s_i \in St(s_k; k \in I_n) \right\}.$$ We define 
\begin{itemize}
	
	\item  $k = \min\{ i \in I_n| s_i \in St(s_j)\}$ and let $p_j$ the endpoint of the horizontal path over $[s_i,s_j]$ starting at $p_i.$   
	\item for $i \in I_n,$ $g_{i,j}$ is defined such as $p_j.g_{i,j}$ is the endpoint of $[s_i,s_j]$ starting at $p_i.$
	\item $I_{n+1} = I_n \cup \{j\}.$
\end{itemize}
	
\end{enumerate}

 The discretization thus describes the holonomy of the connection along the 1-vertices. We have a first sequence $(p_n)_\N$ which stands as a slice of the pull-back $\left((s_n)_\N\right)^* P$ and if $\mathcal{K}_1$ is the 1-skeleton of $\tau,$ the family $(g_{i,j})_{i<j},$ expresses holonomy elements of the connection $\theta$ on the vertices of $\mathcal{K}_1.$ The holonomy of a smooth path $\gamma,$ for a fixed connection, can be approximated by the discretized holonomies computed along a piecewise smooth path along the vertices of the triangulation, close enough to $\gamma.$  In the sequel, we note by $|\sigma|$ the length (resp. $n-$dimensinal the Haussdorf volume) of the 1-vertex (resp. the $n-$simplex or $n-$cube) $\sigma.$
 
 Comparing to Whitney's discretization \cite{Wh}, this scheme does not depend on any exterior trivialization of the principal bundle $P.$ In the sequel, we work with triangulations but the same can be done for cubifications.
 \section{Continuum limits}
\subsection{Differential geometric quantities at the continuum limit} \label{2}
Let us now consider the classical continuum limit scheme. 
Let $(\tau_n)$ be a sequence of triangulations of $M$ such that $\tau_{n} \subset \tau_{n+1}$ (subtriangulations) and such that the length of 1-vertices converge uniformly to $0.$ Then, for fixed $s \in \tau_n \subset... \subset \tau_{n+p} \subset...,$ we have the following 
\begin{itemize}
	\item Let $X$ be a germ of a path $\gamma: t \rightarrow P$ on $P$ at the parameter $t=0$ such that $\pi(\gamma) = \sigma_n$ is a 1-vertex of $\tau_n$ for $n$ large enough, and we assume with no loss of generality in the sequel that $\gamma$ is parametrized by arc-length of $\tau_n.$ Then  $\pi (\gamma_{[0;|\sigma_{n+p}|]})= \sigma_{n+p},$ for $p \in \N,$ and $$\theta(X) = \lim_{p \rightarrow +\infty} \frac{g(p) - I}{|\sigma_p|}$$ where $g_p \in G$ is defined by $\gamma(|\sigma_{n+p}|) = H_{\gamma(0)}\sigma_{n+p}(|\sigma_{n+p}|). g_p \in P.$    
	\item Let $X,Y$ be germs of paths $\gamma_0, \gamma_1 : t \rightarrow P$ on $P$ at the parameter $t=0,$ with $\gamma(0)=\gamma'(0),$ such that there exists a 2-simplex $ \sigma_n \in \tau_n$ for $n$ large enough, where $\gamma$ and $\gamma'$ project on 1-simplexes of $\partial\sigma_n, $ and we also assume arc-length parametrisation as in the previous item. Then $\partial\sigma_{n+p}$ is a piecewise smooth loop parametrized by arc-length, staring from $\pi(\gamma(0))$ along $\pi(\gamma),$ and ending along $\pi(\gamma').$ Then we have $$\Omega(X,Y) =  C\lim_{p \rightarrow +\infty} \frac{Hol_{\gamma(0)}\partial \sigma_{n+p}(|\partial \sigma_{n+p}|) - I}{|\sigma_p|},$$ where $C$ is a constant.
	\item These two items  correspond heuristically to a directionnal derivative of the holonomy at the continuum limit.          
\end{itemize}
We notice here that curvature elements, approximated by loop holonomies, are more easy to define than connection elements. 
\subsection{Measures and cylindrical approximations} \label{2'}

Let us now assume that $G$ is unimodular, equipped with its Haar measure $\lambda_G.$ In the classical appriximation scheme, through Whitney approximation \cite{Wh},  one considers cylindrical approximations via integrable functions defined on the finite dimensional vector spaces spanned by classically discretized connection, see e.g. \cite{AHM,AZ1990}. Passing to the continuum limit, the finite dimensional Lebesgue measure "converge" to the heuristic infinite dimensional Lebesgue measure, see e.g. \cite{AHM,Ma2016-3} for various approaches. 
With this news approximation scheme, we first have to remark that the standard Lebesgue measure on the Lie algebra $\mathfrak g$ does not push-forward to a  
measure on $G$ for e.g. $G= SU(n).$ Conversely, the Haar measure $\lambda_G$ does not generate by pull-back the Lebesgue measure on $\mathfrak g.$ 

\section{Discretized analogs to gauge theories.}\label{3}

\subsection{General principles}
From the previous picture, for a fixed seqence of triangulations $\tau_n$ as before, we can now discretize the continuum gauge theories through the following procedure for action functionnals defined for connections on the principal bundle $P$ over the base $M:$

\subsubsection{Integration over M}
	 Integration over $M$ is performed by summation along the $0-$vertices $$\int_M f = \lim_{n \rightarrow+\infty} \left( Cs_k(Vol(\sigma_{n,k})f(s_{n,k})) \right)$$ where, for fixed $n, $ $(s_{n,k})$ is a family, indexed by $k,$ of nodes (i.e. $0-$vertices) of $\tau_n$ such that each $s_{n,k}$ is a $0-$vertex of $\sigma_{n,k}\subset \tau_n$ and $Cs_k$ denotes the Cesaro mean with respect to the indexes $k.$ Passing from $\tau_n$ to $\tau_{n+1},$ the sequence of nodes $(s_{n,k})_k$ is also an (increasing) subsequence of $(s_{n+1,k})_k.$ The family of nodes must be in a one-to-one correspondence to the family of simplexes of maximal dimension, that is, each $s_{n,k} $ is a node associated to each simplex $\sigma_{n,k}$ of $\tau_n.$ This is an integration analogous to Riemann integration with respect to marked subdivisions. 
\subsubsection{Functions of the curvature} 
A gauge invariant  expression $e^{-i\int_M tr(\Omega^q)}$ or $e^{-\int_M tr(\Omega^q)},$ when $dim M = 2q,$ where $P$ is a gauge-invariant monomial of degree q, can be decomposed decomposed as 
 \begin{eqnarray*} e^{-\int_M tr(\Omega^q)} & = & e^{-\sum_{\sigma \in \tau_n}\int_\sigma tr(\Omega^q)}\\
 	& = & \prod_{\sigma \in \tau_n}e^{-\int_\sigma tr(\Omega^q)}
\end{eqnarray*}
and hereafter there is only one choice to make in order to understand via holonomy the term $\int_\sigma tr(\Omega^q).$ The choice of an adequate gauge for approximating $\Omega$ is not important here, since $tr$ is gauge-invariant. Let $s$ be the node associated to $\sigma \in \tau_{n}.$ The discretization of $\Omega^q$ is related to fixing $\delta_1,...\delta_2q,$ some $1-$vertices of $\sigma$ starting at $s$ such that the oriented simplex $\sigma$ reads as $$\sigma = <\delta_1 , \delta_2 , ... , \delta_{2q}>,$$ which allows to define, setting $X_i$ as the 1-germ of $\delta_i$ at $s,$ 
 \begin{eqnarray*}tr(\Omega^q)(X_1,...X_{2q}) & = & \sum_{p \in \mathfrak{G}_{2q}} \frac{(-1)^{|p|}}{(2q)!} tr\left(\Omega(X_{p(1)},X_{p(2)})...\Omega(X_{p(2q-1)},X_{p(2q)})\right) \\
 	& \sim & \sum_{p \in \mathfrak{G}_{2q}} \frac{(-1)^{|p|}C^q}{(2q)!\prod_{k = 1}^q |<\delta_{p(2k-1)} , \delta_{p(2k)}>|} \\
 	&& tr\big(Hol_{\partial (<\delta_{p(1)} , \delta_{p(2)}>)}(|\partial(< \delta_{p(1)}, \delta_{p(2)}>)|)-Id
 		...\\
 		&&Hol_{\partial (<\delta_{p(2q-1)} , \delta_{p(2q)}>)}(|\partial (<\delta_{p(2q-1)}, \delta_{p(2q)}>)|) - Id \big) 
 \end{eqnarray*} 
\subsubsection{functions of the curvature and the connection}
We consider here only (symmetric gauge-invariant) monomials $ P $ of order $q$ and expressions of the form $tr(\theta \Omega^{q}) $ on the $(2q+1)-$dimensional manifold $M.$ The same kind of expresssions can be given.

\subsection{Case studies}
\subsubsection{Abelian actions}
When $G$ is abelian, the holonomy of a loop $\gamma$ directly reads $exp(\int_{\gamma }\theta)$ and it is gauge invariant. By the way, as for the standard discretization scheme, any action functionnal based on polynomials of the curvature are gauge-invariant.

\subsubsection{2D and 4D-Yang-Mills analog}
The expressions $$YM_{2D} = tr\big(Hol_{\partial (<\delta_{p(1)} , \delta_{p(2)}>)}(|\partial(< \delta_{p(1)}, \delta_{p(2)}>)|) \big)$$  and $$YM_{4D}=tr\left(\left(Hol_{\partial (<\delta_{p(1)} , \delta_{p(2)}>)}(|\partial(< \delta_{p(1)}, \delta_{p(2)}>)|)-Id\right).\left(
Hol_{\partial (<\delta_{p(3)} , \delta_{p(4)}>)}(|\partial (<\delta_{p(3)}, \delta_{p(4)}>)|) - Id \right)\right) $$ are gauge-invariant. Therefore, the discretized gauge group $G^\tau_n$ characterized by the values of the gauge transformations on the nodes act trivially on $YM_{2D}$ and $YM_{4D}.$
By the way any gauge-fixing is rigorous in this modified discretized version of the classical Yang-Mills actions functionals.


\begin{thebibliography}{150}
\bibitem[AHKM2005]{AHM} Albeverio, S.; Hoegh-Krohn, R.; Mazzuchi, S.; \textit{Mathematical theory of Feynman Path Integrals; an introduction} 2nd edition;  Lecture Notes in Mathematics \textbf{523} , Springer (2005)
	\bibitem[AZ1990]{AZ1990} Albeverio, S.; Zegarlinski, B.; Construction of convergent simplicial approximations of quantum fileds on Riemannian manifolds \textit{Comm. Mat. Phys.} \textbf{132} 39-71 (1990)
	\bibitem[Hah2004]{Hah2004} Hahn, A.;
	The Wilson loop observables of Chern-Simons theory on
	$\mathbb{R}^3$
	in axial gauge
	,
	\textit{Comm. Math. Phys.}
	\textbf{248}
	, no. 3, 467?499 (2004)
	
	\bibitem[HCR2015]{HCR2015} Huber, M.; Campagnari, D.; Reinhardt, H.; Vertex functions of Coulomb gauge Yang--Mills theory \textit{Phys. Rev. D} \textbf{91}, 025014 (2015) 
	\bibitem[Lim2012]{Lim2012} Non-abelian gauge theory for Chern-Simons path integral on
	$\mathbb{R}^3$
	, \textit{Journal of Knot
	Theory and its Ramifications}
	\textbf{21}
	 no. 4., articleID1250039 (24p) (2012)
	\bibitem[Ma2017]{Ma2016-3} Magnot, J-P.; The mean value for infinite volume measures, infinite products and heuristic infinite dimensional Lebesgue measures; \textit{J. Math.}Vol. 2017 (2017), Article ID 9853672, 14 pages
	\bibitem[Re1997]{Re1997} Reinhardt, H.; Yang-Mills in axal gauge Phys.Rev. D \textbf{55}  2331-2346 (1997)
	\bibitem[SSSA2000]{SSSA} Sen, S.; Sen, S.; Sexton, J.C.; Adams, D.H.; A geometric discretisation scheme applied to the Abelian Chern-Simons theory \textit{Phys.Rev. E} \textbf{61}  3174-3185 (2000)
	\bibitem[Se2008]{Seng2008}Sengupta, A.N.; Gauge Theory in Two Dimensions: Topological, Geometric and Probabilistic Aspects. Pages 109-129 in\textit{ Stochastic Analysis in Mathematical Physics} edited by Gerard Ben Arous, Ana Bela Cruzeiro, Yves Le Jan, and Jean-Claude Zambrini, published by World Scientific (2008).
	\bibitem[Se2011]{Seng2011} Sengupta, A.N.;
	Yang-Mills in Two Dimensions and Chern-Simons in Three
	, in
\textit{	Chern-
	Simons  Theory:   20  years  after}
	,  Editors  Jorgen  Ellegaard  Anderson,  Hans  U.  Boden,  Atle
	Hahn,  and  Benjamin  Himpel.   AMS/IP  Studies  in  Advanced  Mathematics  (pp.   311-320),
	July 2011.
\bibitem[Wh1957]{Wh} 
Whitney, H., Geometric Integration Theory, Princeton University Press, Princeton, NJ, 1957.
\end{thebibliography}
\end{document}